\documentclass[aps, prd, preprintnumbers, floatfix, showpacs, superscriptaddress, twocolumn]{revtex4}

\usepackage{amsfonts}
\usepackage{amssymb,latexsym}
\usepackage{amsmath,amsbsy}
\usepackage{epsfig,bm}
\usepackage{graphicx}

 \def\cJ{{\cal J}}  \def\cL{{\cal L}}
  \def\cO{{\cal O}} 
   
 \def\cV{{\cal V}}  
 
\def\Tr{\mathop{\rm Tr}}

\def\GeV{{\rm GeV}}
\def\MeV{{\rm MeV}}

\newcommand{\nn}{\nonumber}

\def\nn{\nonumber \\}

\newcommand{\tmfloatcontents}{}
\newlength{\tmfloatwidth}
\newcommand{\tmfloat}[5]{
  \renewcommand{\tmfloatcontents}{#4}
  \setlength{\tmfloatwidth}{\widthof{\tmfloatcontents}+1in}
  \ifthenelse{\equal{#2}{small}}
    {\ifthenelse{\lengthtest{\tmfloatwidth > \linewidth}}
      {\setlength{\tmfloatwidth}{\linewidth}}{}}
    {\setlength{\tmfloatwidth}{\linewidth}}  \begin{minipage}[#1]{\tmfloatwidth}
    \begin{center}
      \tmfloatcontents
      \captionof{#3}{#5}
    \end{center}
  \end{minipage}}

\begin{document}

\preprint{JLAB-THY-08-866}

\title{Pion in the Holographic Model with 5D Yang-Mills Fields}

\author{H.~R.~Grigoryan}
\affiliation{Thomas Jefferson National Accelerator Facility,
             Newport News, VA 23606, USA}
\author{A.~V.~Radyushkin}
\affiliation{Thomas Jefferson National Accelerator Facility,
              Newport News, VA 23606, USA}
\affiliation{Physics Department, Old Dominion University, Norfolk,
             VA 23529, USA}
\affiliation{Laboratory of Theoretical Physics, JINR, Dubna, Russian
             Federation}

\begin{abstract}
We study pion in the holographic model of Hirn and Sanz which contains two Yang-Mills fields defined in the
background of the sliced AdS space. The infrared boundary conditions imposed on these fields generate the
spontaneous breaking of the chiral symmetry down to its vector subgroup.
Within the framework of this model, we get an analytic expression for the pion form factor and a compact result
for its radius. We also extend the holographic model to include Chern-Simons term which is required to reproduce
the appropriate axial anomaly of QCD. As a result, we calculate the anomalous form factor of the pion and predict
its $Q^2$-slope for the kinematics when one of the photons is almost on-shell. We also observe that the anomalous
form factor with one real and one virtual photon is given by the same analytic expression as the electromagnetic
form factor of a charged pion. One of the advantages of the present model is that it does not require an infrared
boundary counterterm to   correctly  reproduce the  anomaly of QCD.
\end{abstract}

 \maketitle


\section{Introduction}

During the last few years applications of gauge/gravity duality \cite{Maldacena:1997re} to hadronic physics
attracted a lot of attention, and  various holographic dual models of QCD were proposed in the literature (see,
e.g., \cite{Polchinski:2002jw}-\cite{Dhar:2007bz}). These models were able to incorporate such essential
properties of QCD as confinement and dynamical chiral symmetry breaking, and also to reproduce many of the static
hadronic observables (decay constants, masses), with values rather  close to the experimental  ones.

In our recent papers  \cite{Grigoryan:2007vg,Grigoryan:2007iy,Grigoryan:2007wn} we developed a formalism that
allows to systematically study the meson form factors within  the holographic ``hard-wall'' approach of
Refs.~\cite{Erlich:2005qh,DaRold:2005zs}. We applied it first to form factors and wave functions of vector mesons
\cite{Grigoryan:2007vg,Grigoryan:2007iy} and then \cite{Grigoryan:2007wn}  to  the pion electromagnetic form
factor. In Ref.~\cite{Grigoryan:2008up}, we extended  the holographic dual model of QCD to incorporate the
anomalous $F_{\gamma^* \gamma^*  \pi^0}(Q_1^2,Q_2^2)$ form factor.

In the present paper, we  consider a holographic model of QCD proposed by Hirn and Sanz \cite{Hirn:2005nr}, with $
SU(2)_L \times SU(2)_R $ Yang-Mills (YM) gauge fields living in the background of sliced five-dimensional (5D) AdS
space. Unlike the approach of Refs.~\cite{Erlich:2005qh,DaRold:2005zs}, this model does not require the existence
of an additional degree of freedom dual to the chiral condensate of four-dimensional (4D) QCD, which spontaneously
breaks the chiral symmetry via the Higgs-like mechanism. Instead, the chiral symmetry breaking down to its vector
subgroup $SU(2)_V$ occurs due to the boundary conditions (b.c.) imposed on the infrared (IR) brane.

At the same time, the global $ SU(2)_L \times SU(2)_R $ symmetry of QCD is generated from the requirement that the
fields vanish on the ultraviolet (UV) boundary. The chiral field $ U(x) $, the phase of which describes the
Nambu-Goldstone bosons, appears in this model as a product of Wilson lines connecting IR and UV branes by the
fifth components of left and right gauge fields.

Since the model of Ref.~\cite{Hirn:2005nr}  incorporates YM fields only, it has a single parameter, the cutoff
scale $z_0$, that determines  the size of  masses $m_\rho, m_{a_1}$, etc., of vector and axial-vector mesons, and
also the pion decay   constant $f_\pi$. Still, many features of this construction are similar to those discussed
in the papers \cite{Erlich:2005qh,Son:2003et,Sakai:2004cn}.

The paper is organized in the following way. We start by outlining, in Section II, the basics of the hard-wall
model of Hirn and Sanz. In particular, we write the form of the 5D action, show how to separate gauge fields into
dynamical and source parts, define the chiral fields as Wilson lines, and demonstrate how the boundary conditions
on the fields break the global symmetry of QCD down to the vector subgroup. We also elaborate on the meaning of
the boundary conditions, and present vector and axial-vector fields in a way helpful for further studies.

In Section III, we calculate and analyze the pion form factor. We show that the form factor can be represented
analytically in terms of the modified Bessel function, and obtain a compact analytic result for the pion charge
radius. We further explore the large-$Q^2$ behavior of the form factor and observe good agreement with
experimental data.

In Section IV, we consider the generalization of the AdS/QCD model that includes isoscalar fields and Chern-Simons
term. Using this extended model we describe the calculation of the  $ F_{\gamma^* \gamma^* \pi^0}(Q_1^2,Q_2^2)$
form factor and express it in terms of the pion wave function and two bulk-to-boundary propagators for the vector
currents describing EM sources. We observe that in case of one real photon, the anomalous form factor of the
neutral pion is identical to the electromagnetic form factor of charged pion. We discuss kinematics with one real
and one virtual photon and calculate the value of the $Q^2$-slope of the form factor. We also investigate the
formal limit of large photon virtualities, and compare these results to those obtained in our earlier paper
\cite{Grigoryan:2007wn}. Finally, we summarize the paper.


\section{Outline of the model}

\subsection{The Setup}

The model of Ref.~~\cite{Hirn:2005nr} is   based on the action
\begin{align}
S_{YM} &= -\frac{1}{4g^2_5}\int d^5 x \sqrt{g} \,\Tr\biggl[L_{MN}L^{MN} + R_{MN}R^{MN} \biggr] \ ,
\end{align}
with the metric
\begin{align}\label{metric}
g_{MN}dx^Mdx^N = \frac{1}{z^2}\left(\eta_{\mu \nu}dx^{\mu}dx^{\nu} - dz^2\right) \ ,
\end{align}
where $ z \in (0,z_0] $, $ \eta_{\mu\nu} = {\rm Diag}(1,-1,-1,-1) $, $ \mu, \nu = (0,1,2,3) $, $ M, N =
(0,1,2,3,z) $,
\begin{align}
A_{MN} &= \partial_{M}A_{N} - \partial_{N}A_{M} - i [A_{M},A_{N}] \ ,
\end{align}
and $ A_{M} = t^a A^a_M $, $ A = \{L,R\} $, ($t^a= \sigma^a/2$, with $\sigma^a$ being  Pauli matrices). The gauge
fields  transform as
\begin{align}\label{gauge trans}
A_{M}(x,z) &\rightarrow g_AA_{M}g_A^{-1}(x,z) + i g_A\partial_{M}g_A^{-1}(x,z) \ ,
\end{align}
where $ g_{A}(x,z) \in SU(2)_{A} $. On the UV brane, the boundary conditions $L_{\mu}(x,0) = \ell_{\mu}(x) $ and $
R_{\mu}(x,0) = r_{\mu}(x) $ are assumed, where $ \ell_{\mu}(x) $ and $ r_{\mu}(x) $ are the sources for the left
and right 4D currents. Vector $ V_{\mu} = (L_{\mu} + R_{\mu})/2 $ and axial-vector $ A_{\mu} = (L_{\mu} -
R_{\mu})/2 $ gauge fields are dual to the vector and axial-vector currents of QCD respectively.
Working in the axial-like gauge, in which $ L_z(x,z) = R_z(x,z) = 0 $, one can write the vector $ \hat{V}_{\mu}$
and the axial-vector $\hat{A}_{\mu}$ fields as
\begin{align}
\hat{V}_{\mu}\left(x, z\right) &\equiv V_{\mu}\left(x,z\right) + \hat{V}_{\mu} \left(x,0\right) \ , \\ \nonumber
\hat{A}_{\mu}\left(x,z\right) &\equiv  A_{\mu}\left(x,z\right) + \alpha\left(z\right)\hat{A}_{\mu}\left(x,0\right)
\ ,
\end{align}
where the  so called ``dynamical'' fields $ V_{\mu}\left(x,z\right) $ and $A_{\mu}\left(x,z\right)$ satisfy the
following b.c.
\begin{align}\label{BC1}
  V_{\mu}(x,0) &= 0 \ , \ \
  A_{\mu}(x,0) = 0
\end{align}
on the UV brane.   However, on the IR brane, the vector field obeys Neumann b.c.
\begin{align}\label{BC1IRV}
 \partial_z V_{\mu}(x,z_0) = 0  \ ,
\end{align}
while both of the axial-vector fields $ A_{\mu}$ and $ \hat{A}_{\mu} $ are required to satisfy Dirichlet b.c.
\begin{align}\label{BC1IRA}
A_{\mu}(x,z_0) = 0 \ , \ \ \hat{A}_{\mu}\left(x,z_0\right) = 0  \ .
\end{align}

As pointed out in Ref.~\cite{Hirn:2005nr} (and will be discussed below), in order to avoid the mixing between the
pion and the axial resonances, the function $ \alpha \left(z\right) $ should satisfy the equation
\begin{eqnarray}
\partial_z \left( \sqrt{g} g^{\mu \nu} g^{zz} \partial_z \alpha(z)\right)  &=& 0 \ .  \label{eoma}
\end{eqnarray}
The following b.c. on the function $ \alpha (z) $
\begin{eqnarray}
  \alpha \left(0\right) = 1 \ , \ \ \alpha \left(z_0\right) = 0 \ ,
\end{eqnarray}
are determined from the b.c. in (\ref{BC1}) and (\ref{BC1IRA}). As a result,
\begin{eqnarray}
  \alpha \left(z\right) & = & 1 - {z^2}/{z_0^2} \ .
\end{eqnarray}
The chiral field
\begin{align}
U(x) = \xi_{R}(x)\xi_{L}^{-1}(x) \ ,
\end{align}
is  built   from  the path-ordered Wilson lines:
\begin{align}\label{Wilson lines}
\xi_{L}(x) &= P  \exp \left\{-i \int_0^{z_0} dz' L_z(x,z') \right\} \ , \\ \nonumber \xi_{R}(x) &= P \exp
\left\{-i \int_0^{z_0} dz' R_z(x,z') \right\}  \ .
\end{align}
With respect to the global chiral transformations, the field $U(x)$ transforms  in the same way as the chiral
field in the non-linear sigma model. Therefore, the pion field is build from a product of Wilson lines extending
from one boundary to the other.

\subsection{Meaning of Boundary Conditions}

The Dirichlet b.c. $ V_{\mu}(x,0) = 0 $ and $A_{\mu}(x,0) = 0$, imposed on the UV brane, are equivalent to $
L_{\mu}(x, 0) = R_{\mu}(x, 0) = 0 $. The latter b.c. are important, since for these the residual gauge invariance
is a global symmetry of 4D QCD. Another significance of these b.c. is that they secure  a finite action at the UV
boundary.  Indeed, since the lagrangian $ \sqrt{g}\Tr F^2 $ is singular at $ z = 0 $, the field strengths have to
vanish there to produce a finite action. We can partially fix the gauge by requiring that
\begin{align}
L_M(x, z \rightarrow 0) = 0 \ , \  \ R_M(x, z \rightarrow 0) = 0  \ .
\end{align}
This gauge choice remains unaltered if we perform additional gauge transformations $g^{\rm res}(x,z \rightarrow
0)$ satisfying the condition
\begin{align}
\partial_{M}g^{\rm res}(x,z \rightarrow 0) = 0 \ ,
\end{align}
which means that $ g^{\rm res}(x,z) $ goes to a  constant matrix $ g_{L,R} \in SU(2)_{L,R} $ at $ z = 0 $. In the
holographic model,  $ (g_L, g_R) \in SU(2)_L \times SU(2)_R $ corresponds to the global chiral symmetry of QCD, at
$ z = 0 $.

The other Dirichlet b.c. $ A_{\mu}(x,z_0) = 0 $ (imposed on the IR brane) breaks gauge invariance in the bulk,
requiring $ L_{\mu}(x,z_0) = R_{\mu}(x,z_0) $, which is equivalent to the condition $ g^{\rm res}_L(x,z = z_0) =
g^{\rm res}_R(x,z = z_0) = h(x)$. The resulting breaking of gauge invariance in the bulk leads to the spontaneous
breaking of the global chiral symmetry on the 4D UV brane down to the vector subgroup. As a consequence, the
Wilson lines  $ \xi_{L,R} $ transform as
\begin{align}
\xi_{L,R} = g_{L,R} \ \xi_{L,R} \ h(x)^{\dagger} \ ,
\end{align}
where $ h(x) \in SU(2)_V $ is a local gauge symmetry on the 4D IR brane.

Finally, the remaining Neumann (gauge invariant)  b.c. $ V_{z\mu}(x,z_0) = \partial_z V_{\mu}(x,z_0) = 0 $ is
required to have a unique solution for the equations of motion. This b.c. was chosen to preserve the vector gauge
invariance, since otherwise the breaking of it may lead to spontaneous breaking of the global vector symmetry of
QCD. However, according to the Vafa-Witten theorem \cite{Vafa:1983tf} this does not occur in QCD.


\subsection{Generalities}

It is useful to define the following 4D fields:
\begin{align}
u_{\mu}\left( x \right) &\equiv  i \left\{ \xi_R^{\dag}\left( \partial_{\mu} - i r_{\mu}
  \right) \xi_R - \xi_L^{\dag} \left( \partial_{\mu} - i \ell_{\mu} \right) \xi_L \right\} \ , \\[10pt] \nonumber
\Gamma_{\mu} \left(x \right) & \equiv \frac{1}{2}\left\{ \xi_R^{\dag}\left( \partial_{\mu} - i r_{\mu} \right)
  \xi_R + \xi^{\dag}_L\left(\partial_{\mu} - i \ell_{\mu} \right) \xi_L\right\} \ .
\end{align}
Notice that
\begin{eqnarray}\label{AMZ}
  L_{z \mu} & = & \tilde{\xi}_L  \left( \partial_z V_{\mu} + \partial_z A_{\mu} -
  \frac{1}{2}  \left( \partial_z \alpha \right) u_{\mu} \right) \tilde{\xi}_L^{\dag},
  \\ \nonumber
  R_{z \mu} & = & \tilde{\xi}_R  \left( \partial_z V_{\mu} - \partial_z A_{\mu} +
  \frac{1}{2}  \left( \partial_z \alpha \right) u_{\mu} \right) \tilde{\xi}_R^{\dag} \ ,
\end{eqnarray}
where
\begin{align}
\tilde{\xi}_{A}(x,z) &= P  \exp \left\{-i \int_z^{z_0} dz' A_z(x,z') \right\} \ .
\end{align}
It is straightforward to see that $\tilde{\xi}_{A}(x,z=0) = \xi_{A}(x) $. The field $ \Gamma_{\mu} \left(x \right)
$ will be used later to define the covariant derivative.

Now, we are in a position to explain the particular  choice of  the field $ \alpha (z)$ described earlier. Indeed,
according to  Eq.~(\ref{AMZ}), the term in the 5D action (\ref{LZRZ}), which describes the mixing of axial-vector
and pion fields, is proportional to the integral
\begin{align}\label{integral}
&\int^{z_0}_0 \frac{dz}{z}(\partial_z A_{\mu})(\partial_z \alpha) = - \int^{z_0}_0
\frac{dz}{z}A_{\mu}\left(\partial_z \frac{1}{z} \, \partial_z\alpha\right) \ .
\end{align}
In order to avoid this mixing, one may impose the EOM (\ref{eoma}) for the field $\alpha(z) $, namely,
\begin{align}\label{seom}
\partial_z \left ( \frac{1}{z}\,  \partial_z\alpha  \right )  = 0  \ .
\end{align}
As a result, the integral in Eq.~(\ref{integral}) vanishes automatically.

It is instructive to observe that the part of the YM action with 4D indices can be written as
\begin{eqnarray}\nonumber
\Tr \left(R^2_{\mu \nu} + L^2_{\mu \nu}\right) = \frac{1}{2}\Tr\left(F^2_{+ \mu \nu} + F^2_{-\mu\nu}\right) \ ,
\end{eqnarray}
where
\begin{align}
  F_{\pm \mu \nu} \left( x, z \right) \equiv \xi_L^{\dag} L_{\mu \nu}
  \xi_L \pm \xi_R^{\dag} R_{\mu \nu} \xi_R \ .
\end{align}
The field strength tensors  of the sources are   defined by
\begin{eqnarray}
f_{\pm \mu \nu} \left( x \right) & \equiv & \xi_L^{\dag}\ell_{\mu \nu} \xi_L \pm \xi_R^{\dag} r_{\mu \nu} \xi_R \
.
\end{eqnarray}
One can rewrite the field $F_{+ \mu \nu}$ as follows
\begin{align}
&F_{+ \mu \nu} = 2 \left( \nabla_{\mu} V_{\nu} - \nabla_{\nu} V_{\mu} - i \left[ V_{\mu}, V_{\nu} \right] - i
\left[ A_{\mu}, A_{\nu}\right] \right) \\ \nonumber  &+ i \alpha \left( \left[ u_{\mu}, A_{\nu} \right] + \left[
A_{\mu}, u_{\nu} \right] \right) + f_{+ \mu \nu} + i \frac{1 - \alpha^2}{2}  \left[u_{\mu},u_{\nu} \right] ,
\end{align}
where the covariant derivatives of the fields are given by  $\nabla_{\mu} \cdot =
\partial_{\mu} \cdot + \left[ \Gamma_{\mu}, \cdot \right]$.
In the same way, the field $F_{- \mu \nu}$ can be rewritten as
\begin{align}
&F_{- \mu \nu} = 2 \left( \nabla_{\mu} A_{\nu} - \nabla_{\nu} A_{\mu} - i \left[ V_{\mu}, A_{\nu} \right] - i
\left[ A_{\mu}, V_{\nu} \right] \right) \nonumber \\[7pt]
&+ i \alpha \left( \left[u_{\mu}, V_{\nu} \right] + \left[V_{\mu}, u_{\nu} \right] \right) +  \alpha f_{- \mu \nu}
\ .
\end{align}
Notice, that the fields $ A_{\mu}$ and $ V_{\mu}$ are dynamical fields (to be discussed in details in the sections
below). These fields contain axial-vector and vector mesons only. Information about the pion field is contained in
the fields $u_{\mu}$ and $\Gamma_{\mu}$.


\subsection{Vector fields}

The dynamical vector fields   have the following representation
\begin{align}
V_{\mu}(x,z) = \sum^{\infty}_{n=1}V^{(n)}_{\mu}(x)\psi_n(z) \ ,
\end{align}
in terms  of the wave functions  $ \psi_n(z) $ satisfying EOM
\begin{align}
\left[z^2 \partial_z^2 - z \partial_z + M^2_n z^2 \right]\psi_n(z) = 0 \ ,
\end{align}
with b.c. $ \psi_n(0) = \partial_z\psi_n(z_0) = 0 $. Here, e.g. the field $ V^{(1)}_{\mu}(x) = g_5\rho_{\mu}(x) $
describes the $\rho$-meson. The solution for $ \psi_n(z) $ is
\begin{equation}
\psi_n(z) = \frac{\sqrt{2}}{z_0 J_1(\gamma_{0,n})}\, z J_1(M_{n} z) \ ,
\end{equation}
where $M_n$ is determined from $ J_0(M_n z_0) = 0 $ and, therefore, $ M_n = \gamma_{0,n}/z_0 $ (with $
J_0(\gamma_{0,n}) = 0 $). The value of $ z_0 = 1/(323 \ \MeV) $ is fixed from the experimental mass of the
$\rho$-meson $ M_1 = 776 \ \MeV $. Eigenfunctions $ \psi_n $ are normalized as
\begin{align}
\int^{z_0}_0~ \, \frac{dz}{z} \, |\psi_n(z)|^2 = 1 \ .
\end{align}

The Fourier transform of the vector field is written as $ V_{\mu}(q,z) = \tilde{V}_{\mu}(q){\cal V} (q,z) $, where
$\tilde{V}_{\mu}(q)$  is the Fourier transform of the 4-dimensional field $V_{\mu} (x)$,  and ${\cal V} (q,z)$ is
the bulk-to-boundary propagator.  The latter satisfies the  EOM
\begin{equation}\label{JQzEOM}
z \,  \partial_z\left(\frac{1}{z}\, \partial_z {\cal V}(q,z)  \right)
 + q^2\,  {\cal V} (q,z) =0
\end{equation}
with b.c. ${\cal V}(q,0) = 1 $ and $ \partial_z{\cal V}(q,z_0) = 0 $. It can be also written as the sum
\begin{equation}
\label{Jmeson}  {\cal V } (q,z) = -g_5\sum_{n = 1}^{\infty}\frac{ f_{n} \psi_n(z)}{q^2 - M^2_{n} } \ ,
\end{equation}
where $ f_n $ is the  decay constant of $n^{\rm th}$ vector meson and
\begin{equation}
f_{n} = \frac{1}{g_5}\left[\frac{1}{z}\partial_z \psi_n(z)\right]_{z=0} = \frac{\sqrt{2}M_n}{g_5 z_0
J_1(\gamma_{0,n})} \ .
\end{equation}

\subsection{Axial-vector and pion fields}

The dynamical axial-vector fields can be written as:
\begin{align}
A_{\mu}(x,z) = \sum^{\infty}_{n=1}A^{(n)}_{\mu}(x)\psi^A_n(z) \ ,
\end{align}
where the functions $ \psi^A_n(z) $ satisfy the same EOM as $ \psi_n(z) $,
 but with different b.c. $ \psi^A_n(0) = \psi^A_n(z_0)
= 0 $. Here, in particular, the field $ A^{(1)}_{\mu}(x) = g_5 a_{1\mu}(x) $ describes $a_1$-meson.

The solution for the axial-vector sector is
\begin{equation}\nn
\psi^A_n(z) \propto z J_1(M^A_{n} z)   \  ,
\end{equation}
where $M^A_n$ is determined from IR b.c.  $ J_1(M_n^A z_0) = 0 $.

In the axial gauge, the axial-vector field with the dynamical fields turned off is  given by
\begin{eqnarray}
\hat{A}_{\mu} \left( x, z \right) &=& \alpha\left( z \right)\hat{A}_{\mu} \left( x, 0\right) \nonumber \\ &=&
\frac{i\alpha\left( z \right)}{2}\left\{\xi_L^{\dag}\partial_{\mu} \xi_L - \xi_R^{\dag}\partial_{\mu}\xi_R
\right\} \ .
\end{eqnarray}
Taking into account the definition of Wilson lines $ \xi_{L,R}(x) $, one can check that
\begin{align}
\xi_L^{\dag}\partial_{\mu} \xi_L = - (\partial_{\mu}\xi_L^{\dag})\xi_L = -i \int_0^{z_0} dz'
\partial_{\mu}L_z(x,z') \ ,
\end{align}
and, therefore,
\begin{eqnarray}\label{defpifield}
\hat{A}^a_{\mu}(x,z) = \alpha(z)\partial_{\mu} \int^{z_0}_0 dz' \ A^a_z(x,z') \equiv \alpha(z)(\partial_{\mu}
\pi^a)
\end{eqnarray}
Notice, that the same result could be obtained even simpler if one uses the additional gauge redundancy by fixing
$ \xi_L^{\dagger} = U $ and $ \xi_R = 1 $,  in which case
\begin{eqnarray}\label{ordpifield}
\hat{A}_{\mu} \left( x, z \right) = \frac{i\alpha\left( z \right)}{2}U\partial_{\mu}U^{\dagger} \ ,
\end{eqnarray}
and since $ U \equiv e^{2i\pi} $, then $ U\partial_{\mu}U^{\dagger} = -2i\partial_{\mu}\pi $, therefore, $
\hat{A}_{\mu} = \alpha(\partial_{\mu} \pi) $.

If $ \ell_{\mu} = r_{\mu} = 0 $ and $ A_{\mu} = V_{\mu} = 0 $, one has
\begin{align}
u_{\mu}\left(x\right) &= i \left\{ \xi_R^{\dag}\partial_{\mu}\xi_R - \xi_L^{\dag}\partial_{\mu}\xi_L \right\} \ ,
\end{align}
\begin{eqnarray}\label{LLRR}
  L_{z \mu} & = & -\frac{1}{2}\left( \partial_z \alpha \right) \xi_L u_{\mu} \xi_L^{\dag} \ ,
  \\ \nonumber
  R_{z \mu} & = & \frac{1}{2}\left( \partial_z \alpha \right) \xi_R u_{\mu} \xi_R^{\dag} \ .
\end{eqnarray}
Notice, that
\begin{align}\label{Uxi}
  \xi_R u_{\mu} \xi_R^{\dag} &= - i U\partial_{\mu}U^{\dagger} \ , \\ \nonumber
  \xi_L u_{\mu} \xi_L^{\dag} &= - i U^{\dagger}\partial_{\mu}U  \ .
\end{align}
The order $ \cO(p^2) $ kinetic term in the action for the chiral fields $U(x)$ is coming from the following part
of the 5D action:
\begin{align}\label{LZRZ}
S_{{\rm kin}} &= \frac{1}{2g^2_5}\int d^4 x \int^{z_0}_0 \frac{dz}{z} \ \Tr\left(L^2_{z\mu}+ R^2_{z\mu} \right) \
.
\end{align}
Taking into account Eqs.~(\ref{LLRR}) and (\ref{Uxi}), and integrating over $ z $, one obtains  that the kinetic
term in the action for the chiral fields becomes
\begin{align}
S_{{\rm kin}} &= \int d^4 x \ \frac{f^2_{\pi}}{4}\Tr\left(\partial_{\mu}U^{\dagger}\partial^{\mu}U \right)  \ ,
\end{align}
where
\begin{align}\label{fpi2}
f^2_{\pi} &= \frac{1}{g^2_5}\int^{z_0}_{0}\frac{dz}{z} \left(\partial_z \alpha\right)^2 = \frac{2}{g^2_5z^2_0} \ .
\end{align}
Furthermore, integrating by parts and using Eq.~(\ref{seom}) gives
\begin{align}
f^2_{\pi} = \frac{1}{g^2_5}\int^{z_0}_{0}\frac{dz}{z} \left(\partial_z \alpha\right)^2 = -
\frac{1}{g^2_5}\left(\frac{\alpha (z)}{z} \
\partial_z \alpha(z)\right)_{z=0} \ .
\end{align}
Since $\alpha (0) $ is  normalized to $1$,  the pion decay constant $f_\pi $ is determined by the value of the
function $\alpha' (z)/ z$ at $z=0$. This result is   similar to that obtained within the holographic model of
Refs.~\cite{Erlich:2005qh,DaRold:2005zs}, where $f_\pi ^2$   is given by the $z=0$ value of the function
\begin{align}
- \frac{1}{g_5^2}  \,  \left ( \frac{1}{z} \,   \partial_z \Psi (z)  \right ) \  ,
\end{align}
with $\Psi (z) $  being   the  pion wave function of that model. As we argued in Ref.~\cite{Grigoryan:2007wn}, it
is the function  $\Phi (z) \sim   \Psi^{\prime}  (z) /z$ that is the most  direct analog of quantum-mechanical
wave functions of bound states. Thus, in the present model we can introduce an analogous  function
\begin{align}
\varphi (z) \equiv - \frac{1}{g_5  f_\pi}  \,  \left ( \frac{1}{z} \,   \partial_z \alpha (z)  \right ) \  ,
\end{align}
which has  $ \varphi (0) =g_5  f_\pi $   normalization at the origin. In fact, given the explicit form of $\alpha
(z)$, one finds that \mbox{$\varphi (z) =g_5  f_\pi $ for all $0 < z \leq z_0$.}

Finally, the full axial-vector field in the axial gauge is:
\begin{eqnarray}
\hat{A}_{\mu}(x,z) = \alpha(z)\partial_{\mu} \pi(x) + \sum^{\infty}_{n=1}A^{(n)}_{\mu}(x)\psi^A_n(z)  \ .
\end{eqnarray}
The longitudinal part of  the axial-vector field $ A^a_{\parallel \mu}(x,z) =
\partial_{\mu}(\pi^a\alpha) $ can be written as
\begin{align}
A^a_{\parallel \mu}(p,z) =  ip_{\mu}\pi^a(p)\alpha(z) \ ,
\end{align}
where $A^a_{\parallel \mu}(p,z)$ and $\pi^a(p)$ are the Fourier transforms of $ A^a_{\parallel \mu}(x,z)$ and
$\pi^a(x) $, respectively. Furthermore, since $A^a_{\parallel \mu}(p,z) = \tilde{A}^a_{\parallel \mu}(p) \,
\alpha(z)$, then
\begin{align}
\pi^a(p)\alpha(z) = -\frac{ip^{\alpha}}{p^2}\tilde{A}^a_{\parallel \alpha}(p)\alpha(z) \  .
\end{align}
This allows us to rewrite $ A^a_{\parallel \mu}(p,z) $ in  the form
\begin{align}
A^a_{\parallel \mu}(p,z) = \frac{p^{\alpha}p_{\mu}}{p^2}\tilde{A}^a_{\parallel \alpha}(p)\, \alpha(z)
\end{align}
involving the longitudinal projector $p^{\alpha}p_{\mu}/{p^2}$  and the pion ``wave function'' $\alpha(z)$.

\subsection{Two-Point Function}

The spectral representation for the two-point function of axial-vector currents can be written  as
\begin{align}
&\langle \, J_{A}^{\alpha}(p)J_{A}^{\beta}(-p)\, \rangle = f^2_{\pi}\frac{p^{\alpha}p^{\beta}}{p^2} \\ \nonumber
&+ \left (- \eta^{\alpha \beta} + \frac{p^{\alpha}p^{\beta}}{p^2}  \right ) \sum_n  \,  \frac{F^2_{A,n}}{p^2 -
M^2_{A,n}}\ + {\rm (nonpole \ terms )} ,
\end{align}
where $ \langle 0|J_A^{\alpha}|\pi(p) \rangle = if_{\pi}p^{\alpha} $ and $ \langle 0|J_A^{\alpha}|A_n(p, s)
\rangle = F_{A,n}  \epsilon^{\alpha}_n (p,s)$, $ F_{A,n} $ correspond to the $n^{\rm th} $ axial-vector meson
decay constant. Finally, as was shown above,  the pion decay constant $f_\pi$ in this  model is given by
\begin{align}
f^2_{\pi} =\frac{2}{g^2_5 z^2_0}  \ .
\end{align}
The hard-wall scale is usually fixed from  fitting the physical mass of the $\rho$-meson, which gives  $ z_0 =
1/(323 \ \MeV) $. The constant  $g_5$  is  fixed from correspondence between AdS/QCD  results and  the asymptotic
behavior of (perturbative) QCD at large $Q^2$, in which case $ g^2_5 = 6\pi^2/N_c$, and, therefore,
\begin{align}
f^2_{\pi} = \frac{N_c}{3\pi^2 z^2_0} \ .
\end{align}
For $ N_c = 3 $, this gives $ f_{\pi} \simeq 102.8 \ \MeV $  instead of $ f^{\rm exp}_{\pi} = 130.7 \ \MeV $.
Since  $ f_{\pi}^2 \sim \cO({N_c})$, one may speculate that the   difference between   the two   values is related
to ${\cal O} (N_c^0)$ corrections to the $f_\pi^2$  prediction of the AdS/QCD model.

\section{Pion  Electromagnetic Form Factor}

Our   next step is to apply the Hirn-Sanz holographic model \cite{Hirn:2005nr}, which does not have  additional
bulk field dual to the chiral condensate,  for   calculation of  the pion form factor. A  brief discussion of this
form factor was given in the original paper \cite{Hirn:2005nr} by Hirn and Sanz. Our goal is to incorporate  the
formalism developed in Ref.~\cite{Grigoryan:2007wn} (see  also \cite{Kwee:2007nq}),  where it was applied to the
pion form factor within the framework of the AdS/QCD model of Refs.~\cite{Erlich:2005qh,DaRold:2005zs}.

\subsection{Three-point function}

To find the pion form factor, we need to consider three-point correlation function between EM current
$J^{el}_\mu(0)$ and two axial currents $ J^a_{5 \alpha} (x_1) , J_{5 \beta}^{a\dagger}(x_2)$
\begin{align} \label{threepoint}
{\cal T}_{\mu \alpha \beta }(p_1,p_2) &=
 \int d^4x_1 \int d^4x_2\ e^{i p_1x_1 - ip_2 x_2} \ \\ \nonumber
&\times \langle 0|{\cal T}J_{5 \beta}^{\dagger}(x_2) J^{\rm el \,  }_\mu(0) J_{5\alpha} (x_1)|0\rangle \ ,
\end{align}
where $p_1$, $p_2$ are  the   corresponding momenta, with  the momentum transfer carried by the EM source being $
q = p_2 - p_1 $ ($ q^2 = - Q^2 < 0 $). The spectral representation for the three-point  function is
\begin{align}
&{\cal T}^{\mu \alpha \beta }(p_1,p_2) = p_1^{\alpha}p_2^{\beta} (p_1+p_2)^\mu
 \frac{f^2_{\pi}\, F_{\pi}(Q^2)}{p_1^2 p_2^2 } \\ \nonumber &+
\sum_{n,m}  {\rm (transverse \ terms) } \ + {\rm (nonpole \ terms) }   ,
\end{align}
where the pion electromagnetic form factor $F_{\pi}(Q^2)$ is defined as
\begin{equation}
\langle\pi(p_1)|J^{el}_\mu(0)|\pi(p_2)\rangle = F_\pi(q^2)(p_1+p_2)_\mu \ .
\end{equation}
According to the prescription of  Ref.~\cite{Grigoryan:2007wn}, the pion form factor can be obtained from the
three-point function using
\begin{align}\label{proj0}
p_{1\alpha} p_{2\beta} {\cal T}^{\mu \alpha \beta }(p_1,p_2) |_{p_1^2=0, p_2^2 =0}  = (p_1+p_2)^\mu {f^2_{\pi}\,
F_{\pi}(Q^2)} \   .
\end{align}

The part of the 5D lagrangian, which may contribute to the pion form factor, is given by
\begin{align}\nonumber
\sqrt{g}\cL_{YM} &= \frac{1}{4g^2_5z}\Tr \left(R^{\mu \nu} R_{\mu \nu} + L^{\mu \nu} L_{\mu \nu} \right) \\
\nonumber  &-
\frac{1}{2g^2_5z}\Tr\left(L^2_{z \mu} + R^2_{z \mu}\right) \\[7pt] \nonumber &\supset
\frac{i}{4g^2_5z}(1-\alpha^2)\Tr\left(V^{\mu\nu}\left[u_{\mu},u_{\nu}\right]\right) \\ \nonumber &-
\frac{1}{4g^2_5z}\left(\partial_z \alpha \right)^2\Tr u^{\mu}u_{\mu}  \ ,
\end{align}
where $ V^{\mu\nu} = \partial^{\mu}V^{\nu} - \partial^{\nu}V^{\mu} $. Taking into account that $ \hat{A}_{\mu
\parallel}^a(x,0) \subset -u^a_{\mu}/2 $ (ignoring the sources $\ell_{\mu}$ and $r_{\mu}$),
one can derive that
\begin{align}\nonumber
\cL_{\rho\pi\pi} \equiv &-\frac{1}{g^2_5z}(1-\alpha^2)\epsilon^{abc} \left(\partial^{\mu} V^{\nu, a}\right)
\hat{A}^b_{\mu
\parallel} \ \hat{A}^c_{\nu \parallel} \\ \nonumber &- \frac{1}{4g^2_5z}\Tr\left(\partial_z \alpha
\right)^2\left(u^{\mu}u_{\mu}\right) \ .
\end{align}
The second term was left ``as is'', since the sources in $ u_{\mu}$, in combination with Wilson lines, will also
contribute to the pion  form factor.

To calculate the 3-point function, we perform first the Fourier transformation,  so that $ \tilde{V}^a_{\mu}(q,z)
= \tilde{V}^a_{\mu}(q)\cV(q,z) $ is an image of $ V^a_{\mu}(x,z)$, and $ \tilde{A}^a_{\mu \parallel}(p,0) $ is an
image of $ \hat{A}^a_{\mu\parallel}(x,0)$. Then, varying the action corresponding to the first term in
$\cL_{\rho\pi\pi}$ with respect to the sources, $ V^a_{\mu}(q)$, $\tilde{A}^b_{\alpha
\parallel}(p_1)$ and $\tilde{A}^c_{\beta\parallel}(-p_2)$, produces the following 3-point function:
\begin{align}\nonumber
&\langle J_{V,a}^{\mu}(q)J_{\parallel A,b}^{\alpha}(p_1) J_{\parallel A,c}^{\beta}(-p_2) \rangle =
i(2\pi)^4\delta^{(4)}(q + p_1 - p_2) \\ \nonumber &\times \,\epsilon_{abc} \
\frac{p_1^{\alpha}p_2^{\beta}}{p^2_1p^2_2}\, (p_1+p_2)^{\mu} \frac{1}{2g^2_5}\, q^2 \int^{z_0}_{\epsilon}
dz~\frac{1}{z}\, {\cal V} (q,z)\, (1-\alpha^2) \ ,
\end{align}
where, anticipating the limit $p^2_1 \rightarrow 0, \, p^2_2 \rightarrow 0$, we took  $ (p_1q) = -(p_2q) = -q^2/2
$ in the numerator. We also took into account that
\begin{align}
\tilde{A}^a_{\mu \parallel}(p,0) = \frac{p^{\alpha}p_{\mu}}{p^2}\tilde{A}^a_{\mu \parallel}(p) \ .
\end{align}

\subsection{Form factor derivation}

Now, representing
\begin{align}
&\langle J_{V,a}^{\mu}(q) J_{\parallel A,b}^{\alpha}(p_1) J_{\parallel A,c}^{\beta}(-p_2) \rangle \\ \nonumber &=
i(2\pi)^4\delta^{(4)}(q + p_1 - p_2)\, \epsilon_{abc}{\cal T}^{\mu \alpha \beta }(p_1,p_2) \
\end{align}
and applying the projection suggested by Eq.~(\ref{proj0}), we   have
\begin{align}\label{PionFF}
f^2_{\pi} F^{(1)}_\pi(Q^2)  &= -\frac{1}{2g^2_5}\, Q^2 \int_{0}^{z_0} \frac{dz}{z}\, {\cal J} (Q,z)\, \left[1
-\alpha^2(z)\right] \ ,
\end{align}
where $ {\cal J}(Q,z) \equiv {\cal V } (iQ,z)$ is   the bulk-to-boundary propagator taken for spacelike momenta
and explicitly given by
\begin{equation}\label{JQz}
{\cal J} (Q,z) = {Qz}\left[K_1(Qz) + I_1(Qz) \frac{K_0(Qz_0)}{I_0(Qz_0)} \right] \ .
\end{equation}
Integrating  by parts Eq.~(\ref{PionFF}) and using  equations of motion both for $ {\cal J} $ and $ \alpha $
produces
\begin{align}\label{FF}
F^{(1)}_\pi(Q^2) &= -1 + \frac{1}{g^2_5 f_\pi^2}\int_{0}^{z_0} dz \, z \, {\cal J} (Q,z) \
\left(\frac{\partial_z\alpha}{z}\right)^2 \ .
\end{align}
Integrating the second term in $ \cL_{\rho\pi\pi} $ with respect to $z$ gives:
\begin{align}\nonumber
-\frac{1}{4g^2_5}\int^{z_0}_0 \frac{dz}{z}\left(\partial_z \alpha \right)^2\Tr \left(u^{\mu}u_{\mu}\right) =
\frac{f^2_{\pi}}{4}\Tr\left[\left(D^{\mu}U \right ) \left (  D_{\mu}U^{\dagger}\right)\right] \ ,
\end{align}
where $ DU = \partial U + i U \ell - i r U $. Expanding $U$ in powers of $\pi$ produces the local term
\begin{align}\nonumber
\frac{f^2_{\pi}}{4}\Tr\left(D^{\mu}U D_{\mu}U^{\dagger}\right) \supset f^2_{\pi}\epsilon^{abc}V^{\mu}_a(x)
\pi_b(x)\partial_{\mu}\pi_c(x)
\end{align}
that compensates ``$-1$'' in Eq.~(\ref{FF}). Namely, performing Fourier transform,   taking into account that
\begin{align}\nonumber
\pi^a(p) = -\frac{ip^{\alpha}}{p^2}A^a_{\alpha \parallel}(p) \ ,
\end{align}
and $V(x) = [\ell(x) + r(x)]/2 $, with further varying the corresponding action, gives the following result  for
the total pion form factor:
\begin{align}\label{totalFF}
F_\pi(Q^2) &=\frac{1}{g^2_5 f_\pi^2}\int_{0}^{z_0} dz \, z \, {\cal J} (Q,z) \
\left(\frac{\partial_z\alpha}{z}\right)^2 \nonumber \\[10pt]
&= \int_{0}^{z_0} dz  \, z \, {\cal J} (Q,z)\,  \varphi^2  (z)  \  .
\end{align}
Using explicit form of $\varphi  (z)$ and incorporating the result $g_5^2 f_{\pi}^2 = {2}/{z^2_0} $ (\ref{fpi2})
for $ f_{\pi} $ we obtain
\begin{align}\label{totalFFz0}
F_\pi(Q^2) &= \frac{2}{z^2_0}\int_{0}^{z_0} dz  \, z \, {\cal J} (Q,z) \ .
\end{align}
Notice that, since ${\cal J} (0,z) = 1$, we have correct normalization for the pion form factor $ F_\pi(0) = 1 $.

\subsection{Results}

Using  EOM for $ \cJ $ in the last equation and integrating further by parts gives
\begin{align}\label{FFanalytic}
F_\pi(Q^2) &= \frac{2}{Q^2z^2_0}\int_{0}^{z_0} dz  \, z^2 \partial_z\left(\frac{1}{z}\partial_z \cJ(Q,z)\right)
\\[7pt] \nonumber &= \frac{4}{Q^2z^2_0}\left[1 - \frac{1}{I_0(Qz_0)}\right] \ .
\end{align}
From the analytic expression for the form factor, it is straightforward to obtain the pion electric charge radius:
\begin{align}
\label{rpi}
 \langle r^2_{\pi} \rangle \equiv - 6\left(\frac{d F_\pi(Q^2)}{dQ^2}\right)_{Q^2=0} = \frac{9z^2_0}{8}  \ .
\end{align}
Taking $z_0 =1/(323\,{\rm MeV})$ gives numerically \mbox{$\langle r^2_{\pi} \rangle \simeq 0.42 \, {\rm fm}^2$,}
which may  be compared with the experimental value $0.45\, {\rm fm}^2 $ \cite{Eidelman:2004wy}.

In the large-$Q^2$ limit, it follows from Eq.~(\ref{FFanalytic}) that
\begin{align} \label{ffnum}
Q^2F_{\pi}(Q^2) \to  \frac{4}{z^2_0}  \simeq 0.42 \, \GeV^2 \  .
\end{align}
It is interesting to note that the highest  Jefferson Lab's experimental points correspond to $Q^2 F_\pi^{\rm exp}
(Q^2) \approx 0.4\,$GeV$^2$, which is very close to the  holographic model result  of Eq.~(\ref{ffnum}) (see also
Fig.1).

\begin{figure}[h]
\includegraphics[width=3.3in]{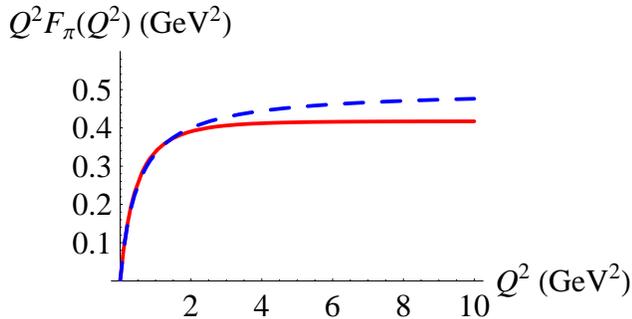}
\caption{\label{FFpi} Pion form factor $Q^2F_\pi(Q^2)$ from the holographic model (solid curve, red online) in
comparison with the monopole data fit $Q^2F_\pi^{\rm mono}(Q^2) = Q^2/(1+Q^2/0.50\,{\rm GeV}^2) $ (dashed curve,
blue online). }
\end{figure}

It is also instructive to use Eq.~(\ref{fpi2}) to substitute $1/z_0^2$ in terms of $f_\pi^2$. This gives
\begin{align} \label{ffnum2}
Q^2F_{\pi}(Q^2) \to  {2g^2_5 f_\pi^2}  \  ,
\end{align}
the  expression  analytically coinciding with our  result obtained in Ref.~\cite{Grigoryan:2007wn} within the
framework of the AdS/QCD model of Refs.~\cite{Erlich:2005qh,DaRold:2005zs}.

This outcome  has very basic  reasons. Namely, the large-$Q^2$  behavior of the form factor is determined, first,
by the large-$Q^2$  form of the bulk-to-boundary propagator ${\cal J} (Q,z)$  (which coincides with its free-field
version ${\cal K} (Qz) \equiv Qz K_1 (Qz) $  in any model) and, second,   by the value of the pion wave function
$\varphi (z)$ (or $\Phi (z)$) at the origin. The latter   also determines  $f_\pi$    in both holographic models,
which results in the same analytic result  $2g^2_5 f_\pi^2$  for the large-$Q^2$ limit  of  $Q^2F_{\pi}(Q^2)$ when
it is expressed in terms of $f_\pi$. However, as we already discussed in Ref.~\cite{Grigoryan:2007wn}, if we take
the experimental value $f_\pi^{\rm exp}  \simeq 131\,$MeV, Eq.~(\ref{ffnum2}) gives the value $0.67 \, \GeV^2$
that is well above Jefferson Lab's experimental points. In this sense, the expression (\ref{ffnum})    for the
limit of $Q^2F_{\pi}(Q^2)$ in terms of $z_0$ is  numerically   more successful than Eq.~(\ref{ffnum2}). One may
speculate that since the pion form factor  in our calculation is given by the ratio of the 3-point function term
(which is proportional to $f_\pi^2 F_\pi (Q^2)$) to the 2-point function term  (which is proportional to
$f_\pi^2$), the overall error  of the model  in the value of $f_\pi$  is cancelled, and the remaining expression
for $F_\pi (Q^2)$ in terms  of $z_0$  correctly reflects information about the pion size.

From the form factor  expression (\ref{totalFF})   and the decomposition of ${\cal J}(Q,z)$ over the $Q$-channel
bound states, one can extract the $ \rho \pi \pi $ coupling:
\begin{align}
g_{\rho \pi \pi} &\equiv \frac{1}{f_{\rho}}\lim_{Q^2 \rightarrow -M^2_1}(Q^2 + M^2_1)F_\pi(Q^2) \\ \nonumber &=
\frac{4g_5}{J_1(\gamma_{0,1})}\int_{0}^{1} dx  \, x^2 \, J_1(\gamma_{0,1} x) \ .
\end{align}
Here we took into account the correct normalization of currents, giving a factor of $ \sqrt{2} $. From this result
it follows that numerically $ g_{\rho \pi \pi}  = 1.383 \, g_5 $. Taking $ g_5 = \sqrt{2}\pi $ we obtain  $
g_{\rho \pi \pi} \simeq 6.15 $. The experimental value  is  $ g_{\rho \pi \pi} \simeq 6.03 \pm 0.07 $. The numbers
obtained in Models A and B of Ref.~\cite{Erlich:2005qh} are $ 4.48 $ and $ 5.29 $ respectively.

\section{Anomaly}

\subsection{Chern-Simons action}

Since the Chern-Simons/Wess-Zumino-Witten term for $ SU(2) $ gauge/global group is vanishing, we need to extend
the flavor symmetry to $U(2)_L \times U(2)_R $, so that the fields are written as
\begin{align}
{\cal B}_{\mu} = t^a B^a_\mu + \frac{1}{2}\hat {B}_\mu \  .
\end{align}
In order not to confuse the hats on the $U(1)$ fields with the hats on the gauge fields in the axial gauge, we
will assign hats only to the former and omit these for the latter.

Extending the holographic dictionary, the 4D isosinglet vector $ J_{\mu}^{\{I=0\}}(x) $ current will correspond to
\begin{align}
J_{\mu}^{\{I=0\}} &= \frac{1}{2}\left (\bar{u} \gamma_{\mu} u + \bar{d} \gamma_{\mu} d \right)= \frac{1}{2}\bar{q}
\gamma_{\mu}\textbf{1}q \rightarrow  \hat {V}_{\mu}(x,z) \ ,
\end{align}
where $ \hat {V}_{\mu}(x,z) \in U(1)_V $ is the  abelian part of the $ U(2)_V $ field. We also remind that the
third component of the isovector $J_{\mu}^{\{I=1\}, a}(x) $ current corresponds to
\begin{align}
J_{\mu}^{\{I=1\}, 3}&= \frac{1}{2}\left(\bar{u} \gamma_{\mu} u - \bar{d} \gamma_{\mu} d \right) = \bar{q}
\gamma_{\mu}\frac{\sigma^3}{2}q \rightarrow  V_{\mu}^3(x,z) \ .
\end{align}
Note  that the EM current of QCD is defined as
\begin{align}
J^{{\rm EM}}_{\mu} = J_{\mu}^{\{I=1\}, 3}+ \frac{1}{3} \, J_{\mu}^{\{I=0\}} \ .
\end{align}
It  has both isovector (``$\rho$-type'') and isosinglet (``$\omega$-type'') terms.

The $ {\cal O} (B^3) $ part of the 5D CS action, in the axial gauge $B_z = 0$ is
\begin{align}\nonumber
S^{(3)}_{\rm CS}[{\cal B}] = \frac{N_c}{24\pi^2}\epsilon^{\mu\nu\rho\sigma}& {\rm Tr} \int d^4 x\, dz
\left(\partial_z {\cal B}_{\mu}\right)\biggl[{\cal F}_{\nu \rho}{\cal B}_{\sigma} + {\cal B}_{\nu}{\cal F}_{\rho
\sigma} \biggr] \ ,
\end{align}
where $ {\cal F}_{\mu \nu} = \partial_{\mu}{\cal B}_{\nu} - \partial_{\nu} {\cal B}_{\mu}  $. In the holographic
model (cf. \cite{Domokos:2007kt}),  the CS term is
\begin{align}
S^{\rm AdS}_{\rm CS}[{\cal B}_L, {\cal B}_R] = S^{(3)}_{\rm CS}[{\cal B}_L]  - S^{(3)}_{\rm CS}[{\cal B}_R]  \ ,
\end{align}
where $ {\cal B}_{L,R} = {\cal V} \pm {\cal A} $ and $ {\cal V}({\cal A}) \in U(2)_{V(A)}$.

After long,  but straightforward calculations, we get
\begin{align}
S^{\rm AdS}_{\rm CS} &= \frac{N_c}{12\pi^2}\epsilon^{\mu\nu\rho\sigma} \int d^4 x \int_0^{z_0} dz \ \pi^a \\
\nonumber &\times \left[ 2\left(\partial_z \alpha\right)\left(\partial_\rho V^a_{\mu} \right)
\left(\partial_{\sigma}\hat{V}_{\nu}\right) - \alpha \partial_z\left(\partial_\rho V^a_{\mu} \
\partial_{\sigma}\hat{V}_{\nu}\right)\right] \ .
\end{align}
Integrating the second term by parts with respect to $ z $ and taking appropriate care on the  IR boundary gives
\begin{eqnarray}\nonumber
S^{\rm AdS}_{\rm CS} = \frac{N_c}{4\pi^2}\epsilon^{\mu\nu\rho\sigma}\int_0^{z_0} dz \, (\partial_z \alpha ) \int
d^4 x \ \pi^a \left( \partial_\rho  V^a_{\mu} \right) \left(\partial_{\sigma}\hat{V}_{\nu}\right) \\
\label{grhopi}
\end{eqnarray}
Recall that $ \alpha(z) = 1 - z^2/z^2_0 $ and, in this  model, it has the meaning of the pion ``wave function''.

\subsection{Anomalous Form Factor}

In QCD,  the $\gamma^* \gamma^* \pi^0 $ form factor is defined  by
\begin{align}
\int d^4 x \ e^{-iq_1 x }& \langle {\pi}, {p}|T\left\{J^{\mu }_{\rm EM}(x)\,J^{\nu}_{\rm EM}(0)\right\}| 0 \rangle
\\ \nonumber &= \epsilon^{\mu  \nu \alpha  \beta}q_{1 \, \alpha} q_{2\, \beta} \, F_{\gamma^*\gamma^*\pi^0} \left(Q_1^2,Q_2^2
\right ) \ ,
\end{align}
where $ p = q_1 + q_2 $  is the pion momentum, $ q_{1}, q_{2} $ are the momenta of photons, and  $ q^2_{1,2} =
-Q^2_{1,2} $.

Varying $ S^{\rm AdS}_{\rm CS} $ gives the 3-point function:
\begin{align}
T_{\alpha \mu \nu}(p,q_1,q_2) = & \frac{N_c}{12 \pi^2}\frac{p_{\alpha}}{p^2} \, \epsilon_{ \mu \nu  \rho \sigma}
\, q_{1}^\rho q_{2}^{\sigma}K(Q^2_1,Q^2_2)
\end{align}
with
\begin{eqnarray} \label{K12}
K(Q_1^2,Q_2^2)  = -\int_0^{z_0}  {\cal J}(Q_1,z) {\cal J}(Q_2,z)\,   \partial_z \alpha(z) \, dz \ ,
\end{eqnarray}
where $\cJ(Q,z)$ is the non-normalizable mode, see Eq.~(\ref{JQz}). It satisfies EOM given by Eq.~(\ref{JQzEOM}),
and is normalized by  $ \cJ(0,z) = 1 $ for $Q=0$. Note  that deriving the result for $K(Q_1^2,Q_2^2)$ we used the
fact that both isoscalar and isosinglet vector mesons are described by the same EOM and b.c.

QCD axial anomaly requires: $ K^{QCD}(0,0) = 1 $. Indeed, in the present extended holographic model, we get:
\begin{align}
K(0,0) = - \int_0^{z_0}\partial_z \alpha(z) \, dz  = \alpha(0) = 1  \ .
\end{align}
This result for the anomalous form factor is very similar to that  obtained in our paper \cite{Grigoryan:2008up},
where we  worked out a CS  extension of the hard-wall model of Refs.~\cite{Erlich:2005qh,DaRold:2005zs}. However,
in  the present model, we do not have a bulk field dual to the chiral condensate operator of QCD, and, moreover,
we do not need to add a counterterm on the IR boundary to reproduce the correct normalization of the anomalous
form factor.

When  only one of the photons is   virtual, $ Q^2_1 = Q^2 $,  while  another is real, $Q^2_2 = 0 $, we have
\begin{eqnarray} \label{KQ0}
K(Q^2,0)  = \frac{2}{z^2_0}\int_0^{z_0} z \ {\cal J}(Q,z) \, dz \ .
\end{eqnarray}
It is  easy to notice that this expression for $K(Q^2,0) $ coincides with the expression (\ref{totalFFz0}) for
$F_\pi (Q^2)$, i.e., the anomalous form factor $K(Q^2,0)$ in this model coincides with the pion EM form factor!

The slope $ a_{\pi} $ of the anomalous form factor  defined as
\begin{align}
a_{\pi} &= -m^2_{\pi}\left[\frac{d K(Q^2,0)}{dQ^2}\right]_{Q^2=0}  \  ,
\end{align}
in the present model is  given by
\begin{align}
a_{\pi} & = \frac{3}{16}m^2_{\pi}z^2_0\ .
\end{align}
Numerically, we have $ a_{\pi} \simeq 0.035 $ (compare with the recent result in Ref.~\cite{Pomarol:2008aa}). This
number is not very far from the central values  of two last experiments, $a_\pi =0.026 \pm 0.024 \pm 0.0048$
\cite{Farzanpay:1992pz}, $a_\pi =0.025 \pm 0.014 \pm 0.026$ \cite{MeijerDrees:1992qb}, but the experimental errors
are rather  large. An earlier experiment \cite{Fonvieille:1989kj} produced $a_\pi = -0.11  \pm 0.03 \pm 0.08$, a
result whose central value has opposite sign and much larger absolute magnitude. In  the spacelike   region,  the
data are available only for  the values \mbox{$Q^2\gtrsim 0.5$\, GeV$^2$}  (CELLO~\cite{Behrend:1990sr}) and
\mbox{$Q^2\gtrsim 1.5$\, GeV$^2$} (CLEO~\cite{Gronberg:1997fj}) which cannot  be treated as very small. The CELLO
collaboration~\cite{Behrend:1990sr}   gives   the value   \mbox{$a_\pi = 0.0326\pm 0.0026$} that   is   very close
to our result. To settle the uncertainty of the timelike data  (and also on its own grounds), it would be
interesting to have data on the slope from the spacelike   region of very   small $Q^2$, which may be obtained by
modification of the PRIMEX experiment \cite{primex} at JLab.

\subsection{$\rho \omega \pi$ coupling}

Substituting $\hat{V}_{\nu}(x,z) = g_5 \omega_{\nu}(x)\psi_1(z) $ and $V^a_{\mu}(x,z) = g_5
\rho^a_{\mu}(x)\psi_1(z) $ into Eq.~(\ref{grhopi}), we  obtain
\begin{eqnarray}\nonumber
\cL^{\rm AdS}_{\rho\omega\pi} = -\left[\frac{N_c g^2_5}{2\pi^2 z^2_0f_{\pi}}\int_0^{z_0} dz \, z \ \psi^2_1\right]
\epsilon^{\mu\nu\rho\sigma} \Pi^a \left( \partial_\rho \rho^a_{\mu}
\right)\left(\partial_{\sigma}\omega_{\nu}\right) \ .
\end{eqnarray}
Here,  we   introduced the dimensionful pion field  $ \Pi^a  = f_{\pi} \pi^a $. This lagrangian is similar to that
obtained in the hidden local symmetries approach \cite{Fujiwara:1984mp} (see also a review
\cite{Meissner:1987ge}). Thus, we may write  that
\begin{align}
g_{\rho \omega \pi} &= -\frac{N_c g^2_5}{2\pi^2} \frac{M_{\rho}}{z^2_0f_{\pi}} \int_0^{z_0} dz \, z \ \psi^2_1 \\
\nonumber &= -\frac{N_c g^2_5}{\pi^2} \frac{M_{\rho}}{f_{\pi}}\int_0^{1} dx \, x^3
\frac{J^2_1(\gamma_{0,1}x)}{J^2_1(\gamma_{0,1})} \ .
\end{align}
The numerical value of this coupling is $ g_{\rho \omega \pi}  \simeq -12.77 $ (for $ g_5 = \sqrt{2}\pi $). The
value of this coupling and especially its sign have important phenomenological implementations, see e.g.
Ref.~\cite{Nakayama:2006ps}. However, one cannot directly measure this coupling constant, since the decay $ \omega
\rightarrow \rho \pi $ is energetically forbidden.

\subsection{Large-$Q^2$ behavior}

Equation  (\ref{KQ0})  formally gives prediction for the $K(Q^2,0)$ form factor at all $Q^2$, and it is
interesting to compare it with the monopole fit $K^{\rm CLEO} (Q^2,0) =1 /(1+Q^2/\Lambda_\pi^2)$ (fitted   value
is $\Lambda_\pi= 776$ MeV) of CLEO data  \cite{Gronberg:1997fj} which extend to $Q^2 \sim 10\,$GeV$^2$. The
comparison demonstrating a rather good agreement is shown in Fig.~2, where the value $z_0=1/(323\,{\rm MeV})$ is
taken for the AdS/QCD curve.

\begin{figure}[h]
\includegraphics[width=3.3in]{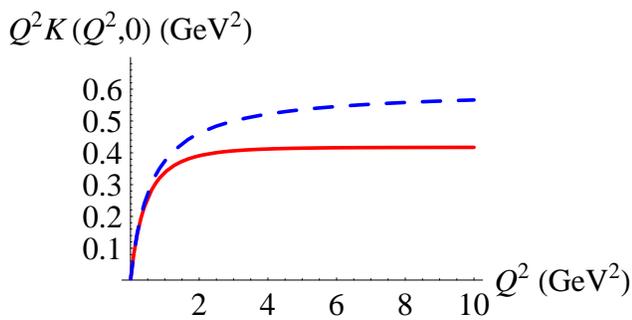}
\caption{\label{FFggpi} Anomalous form factor $Q^2 K(Q^2,0)$ from the holographic model (solid curve, red online)
in comparison with the monopole fit $Q^2 K^{\rm mono}(Q^2,0) = Q^2/(1+Q^2/0.60\,{\rm GeV}^2) $ of CLEO data
(dashed curve, blue online).}
\end{figure}

One may also consider the   general case of large virtualities, when   \mbox{$Q_1^2 = (1+\omega)Q^2$} and  $Q_2^2
= (1-\omega)Q^2$, with $\omega$ being  a  fixed parameter, $-1 \leq \omega \leq 1$ and $Q^2$ large. In this limit,
the bulk-to-boundary propagators  ${\cal J}(Q_1,z),  {\cal J}(Q_2,z)$ in Eq.~(\ref{K12}) may be substituted by
their free-field version ${\cal K} (Qz) = zQ K_1 (Qz)$. Since ${\cal K} (Qz) $  exponentially $\sim e^{-Qz}$
decreases for large $z$, the $z$-integral is   dominated by small $z\sim 1/Q$,   and Eq.~(\ref{K12}) converts into
\begin{align}\label{kQ1Q2}
&K((1+\omega)Q^2,(1-\omega)Q^2)   \\ \nonumber &\to \frac{2}{z_0^2 Q^2} \, \sqrt{1-\omega^2} \int_0^\infty d\chi
\, \chi^3 \,  K_1 ( \chi \sqrt{1+\omega}) K_1 ( \chi \sqrt{1-\omega})  \ .
\end{align}
Using Eq.~(\ref{fpi2}) to substitute the overall factor $2/z_0^2$ by $g_5^2 f_\pi^2$, we obtain exactly the result
derived in our earlier paper \cite{Grigoryan:2007wn}  within the extension of the hard-wall model of
Refs.~\cite{Erlich:2005qh,DaRold:2005zs}. This coincidence in analytic form is   analogous to that observed in the
pion electromagnetic form factor case. Indeed, the large-$Q^2$ asymptotics is determined by the large-$Q^2$
behavior of the bulk-to-boundary propagators, which is the same (free-field-like)  in all the models, and by the
value of the pion wave function $\varphi (0)$  at the origin, which is fixed by the pion decay   constant.
Numerically, however, the results  of  the present model  based on (\ref{kQ1Q2}) would depend on whether one takes
the experimental value for $f_\pi$ or substitutes $g_5^2 f_\pi^2$ by $2/z_0^2$. Again, the situation is the same
as in the pion EM form factor case.

\section{Summary and conclusions}

Working within the framework of the holographic dual model of QCD proposed by Hirn and Sanz, we study form factors
of the pion, namely, the electromagnetic form factor of charged pions and the anomalous form factor $\gamma^*
\gamma^* \pi^0 $ of the neutral pion. In order to calculate the latter, we extend the Hirn-Sanz model by
incorporating the Chern-Simons term into the original 5D action.

Due to a simple form of the pion ``wave function'' $\alpha(z)$, the pion form factor can be written in explicit
analytic form involving the modified Bessel function. This analytic expression gives a simple formula for the pion
charge radius in terms of the hard-wall scale $z_0$, which is  the only parameter of the model fixed by the value
of the $\rho$-meson  mass. Written as a function of $z_0$ and $Q^2$, the prediction of the present model is in a
good agreement with experiment both for low and high $Q^2$ values. We also established that the low energy
coupling constant $g_{\rho\pi\pi}$ in the present model is in better agreement with experiment than the result of
the hard-wall AdS/QCD model of Ref.~\cite{Erlich:2005qh}.

We extended the model  of Hirn  and Sanz by adding the Chern-Simons term and demonstrated that such an extension
correctly reproduces QCD anomaly. There was no need to introduce an IR counterterm which was required in our
previous paper \cite{Grigoryan:2008up}, where we worked in the framework of the hard-wall AdS/QCD model of
Refs.~\cite{Erlich:2005qh,DaRold:2005zs}. We also observed that the anomalous pion form factor with one real and
one virtual photon with momentum transfer $Q^2$ in the present model is given by exactly the same analytic
expression as the form factor of the charged pion evaluated for the same $Q^2$. This outcome may be partially due
to a very simple form of the pion wave function.

We calculated the $Q^2$-slope of the anomalous form factor predicted by the present  model, which was considered
earlier in our paper \cite{Grigoryan:2008up} and was also recently discussed in the Ref.~\cite{Pomarol:2008aa}. In
addition, we calculated the value of the $g_{\rho \omega \pi}$ coupling, which is important for  phenomenological
considerations. Finally, we showed that in the large $Q^2$-region we reproduce the same results as in case of the
hard-wall model of Refs.~\cite{Erlich:2005qh,DaRold:2005zs}.

It is encouraging to establish that such a simple model containing just one free parameter, the confinement scale
$z_0$, produces the results which are in  good agreement with  experimental findings. Moreover, most of the
important expressions can be represented analytically without making any approximations. One can think that the
role of the scalar field in the hard-wall AdS/QCD model is now played by the appropriate b.c. on the IR. However,
with this simplicity, we loose information about the chiral condensate and the dependence of the observables on
it. This dependence was studied in our earlier papers \cite{Grigoryan:2007wn, Grigoryan:2008up}.

\vspace{1cm}
\newpage

\section{Acknowledgments}

H.R.G. would like to thank J.~Erlich, C.~Carone, T.~S.~Lee and C.~D.~Roberts for valuable comments, and
A.~W.~Thomas for support at Jefferson Laboratory.

We thank  the organizers of the program ``From Strings to Things: String Theory Methods in QCD and Hadron
Physics'' at the  Institute for Nuclear Theory  at the University of Washington for support during this   program,
the participation in which  stimulated the completion of this  work.

This paper is authored by Jefferson Science Associates, LLC under U.S. DOE Contract No. DE-AC05-06OR23177. The
U.S. Government retains a non-exclusive, paid-up, irrevocable, world-wide license to publish or reproduce this
manuscript for U.S. Government purposes.

\end{document}